\documentclass[final,5p,times,twocolumn]{elsarticle}
\usepackage{subcaption} 
\usepackage{multirow}
\usepackage{sidecap} 
 \usepackage{booktabs}
\usepackage{amsmath}
\usepackage{subcaption}
\usepackage{epstopdf}
\usepackage{lineno}
\modulolinenumbers[200]

\journal{Microelectronic Engineering}

\begin{document}

\begin{frontmatter}
\title{ Influence of moisture on the operation of a mono-crystalline based silicon photovoltaic cell: A numerical study using SCAPS 1 D}

\author[uy1,aimscm]{ Bayawa MOHAMED\corref{cor1}}
\ead{bayawa.mohamed@aims-cameroon.org}


\author[aims,e2i]{Serges ZAMBOU\corref{cor1}}
\ead{serges@aims.ac.za}

\author[uy1]{ Serge Sylvain ZEKENG}

\cortext[cor1]{Corresponding authors}
\address[uy1]{Laboratory of Materials Science, Department of Physics, Faculty of Science, University of Yaounde 1, Po.Box 812, Yaounde, Cameroon}
\address[aimscm]{African Institute for Mathematical Sciences-Cameroon,Crystal Gardens PO. Box 608 Limbe , Cameroon}
\address[aims]{ African Institute for Mathematical Sciences-South Africa,  5-8 Melrose Road, Muizenberg 7945, Cape Town, South Africa}
\address[e2i]{Ecole d'Ingenierie Industrielle, BP : 9381, Bagangte, Cameroun}


\begin{abstract}
Moisture in the form of humidity has a significant impact on the operation of photovoltaic (PV) cells and panels. Moisture is often diffuse through the encapsulant film (Ethylene Vynil Acetate (EVA), Poly Vinyl Butyral (PVB), Poly Vinyl Fluoride (PVF), Polymethyl Methacrylate (PMMA)) or the breathable back-sheet. Moisture is known to accelerate the degradation and reduce the performance of a PV while operating in a wet environment. Due to the high cost and the time needed to performed such studies, SCAPS 1 D is widely used to simulate and estimate the operation of PV. In this work, we showed the effect of moisture on the operation and performance of a PV, made from monocrystalline silicon. Namely, the influence of defects and impurities generated by moisture was investigated. We studied the simultaneous effects of Silanlols (Si-OH), Hydrogen ions H$^+$, and  the metallic ions generated from the corrosion of contacts, encapsulant and transparent conducting oxide (TCO) when there is production of acid (Al$^{3+}$; Zn$^{2+}$), or from dust (Fe$^{3+}$).

The numerical simulation showed that : starting with PV cells without moisture, the Fill Factor (FF) and the Power Conversion Efficiency (PCE) drop respectively from 82.80 $\%$ to 81.78 $\%$ and 18.57 $\%$ to 11.46 $\%$ , when there is moisture in the cell and PVB used as encapsulant.
Further degradation of parameters was observed, when the moisture leads to the production of acetic acid on the EVA, or when there is dust having iron (Fe). FF and PCE were further dropped to  79.20 $\%$ and 7.04 $\%$ respectively, using the same initial photovoltaic cells. Furthermore,  a net decrease in key electrical parameters of the PV was observed throughout the study,  with maximal power ( P$_{max}$), the short circuit current ( J$_{SC}$), the current density at maximum power  J$_{MP}$ reducing by more than 50\%.
This study paves the way for the improvement of performance, and the understanding of degradation process in panels used in wet environments and tropical area.
\end{abstract}


\end{frontmatter}

\linenumbers

\section{Introduction}

The search for cleaner, sustainable and environmentally friendly power sources has led to the ubiquitous development of PV module in recent decades \citep{baya1,baya2}. Despite the overall rapid improvement of PV performance worldwide, their use is not evenly spread across the globe, due to their relatively high cost per kWh 
compared to other power sources such as hydroelectric, nuclear, coal and wind \citep{baya3}. As a matter of fact, developing country mostly located in tropical regions, with regular and annual exposure to sunlight, have the lowest penetration of PV technology, despite the high need and the low connection of their citizens to the power grid. It is suggested that a tipping point for the PV technology will be reached when the grid parity will be attained, this means the state when the average cost of a PV module will be equivalent to that of the grid for a given geographical region \citep{baya4,baya5}.  In order to achieve that tipping point, Solar cell properties and parameters such as encapsulant, PV materials, desiccant are constantly improving. Moisture often causes an urge challenge in the long-term use of PV in tropical,  humid and hot environments. Despite the development made in encapsulating PV module, they still suffer from moisture ingress \citep{baya6,baya7}. Moisture is accentuated in regions with hot and humid climate, and water diffuses faster into the encapsulant film ( Ethylene Vynil Acetate (EVA) or Poly Vinyl Butyral (PVB)) region. It is showed that the rate of PV degradation can be correlated with its quantity of moisture \citep{baya13}. Moisture is known to accelerate the creation of delamination, back-sheet debonding, obscuration of the encapsulating glass, current leakages, oxidation of the metallic contacts. Moisture also accelerates corrosion and affects the series resistance of the junction, create more ingress paths, lead to the reduction of power and ultimately to the failure of the module \citep{baya8,baya9,baya10,baya11}. 
The practical determination of the PV module degradation due to moisture is usually done according to the IEC 61215 standard, usually via dump-heat or using solar tester module. The degradation due to moisture is made on real modules, with expensive equipment, often time-consuming and requiring a certain level of expertise to be completed. 

The advent of electronics simulation tools about half a century ago, marked a turning point in the field of materials and devices since they significantly contribute to cost saving, devices improvement and optimization of parameters in various aspect of the manufacturing process, operation and rating. In the years 2000, Burgelman et Al. Launch the development of now a very well acclaimed Solar Cell Capacitance Simulator (SCAPS)\citep{baya13}. SCAPS is a one-directional based windows simulation program developed for cell structures of the CIGS, CuInSe$_{2}$, CdTe, crystalline and amorphous Si and GaAs family \citep{baya14}.  Despite the fast development and the improvement of perovskite, and  CIGS solar cell efficiency, mono and polycrystal silicon solar cells are still the most use to date in the production of PV, thus the simulation of their ageing or operation alteration due to moisture is well-deserved. SCAPS offers a wide variety of parameters that can be investigated in dark, light, and variables temperature, this includes the fill factor (FF\%), the quantum efficiency (QE\%), capacitance-voltage spectroscopy C(V), open circuit voltage (V$_{OC}$), short circuit current density (J$_{SC}$), capacitance frequency spectroscopy C($f$), current densities etc \citep{baya15}.

In this paper, SCAPS is used to simulate the effect of moisture on a monocrystalline Silicon-based PV. In that regards, the electrical behaviour of the semiconducting material will be investigated once the moisture/water reached the material. The state of the art parameters will be used in this study, the quantum efficiency, maximal power, current density and other key electrical parameters would be studied. The chemical mechanism of moisture production and metallic elements would be discussed in order to clearly explain the degradation of PV expose to humidity or used in tropical climate. This study will provide an insight of the moisture effect on the electrical and operational behaviour of the solar cell under a prolonged variation of humidity.

\section{Chemical consideration and moisture modeling}

\begin{figure}[!htbp]
\begin{center}
\includegraphics[scale=0.5]{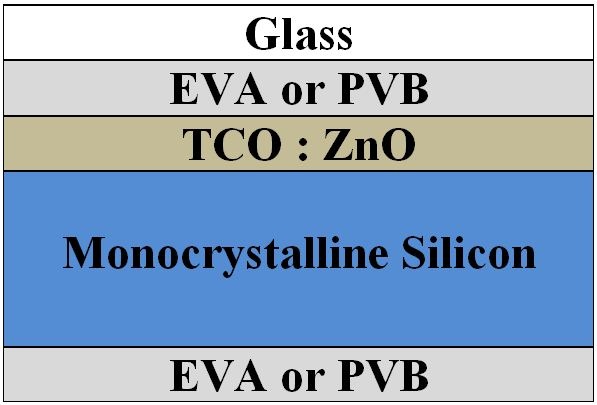}
\end{center}
\caption{Illustration of a basic photovoltaic module. This module is made of several layers from top to bottom : Glass, the encapsulant film (EVA or PVB), the transparent conducting oxide  (TCO) or  (ZnO), the monocrystalline silicon and a PVB or EVA layer.}
\label{figure1}
\end{figure}

For this study, a simplified model of a photovoltaic module was made, the essential layers needed for the efficient and optimal operation of the PV was considered. 5 layers module was chosen as shown in figure \ref{figure1}. The top layer of a PV is made of a glass, which serve as a protection for other materials against factors including radiation, water, impact, in such a way that their efficiency and operation is not altered quickly. In our simulation, we considered the widely use two types of encapsulants, either the Polyvinyl Butyral (PVB) or the Ethylene Vinyl Acetate (EVA). The solar cell was considered in this study to be made of  (TCO) on zinc oxide (ZnO) forming a junction with silicon (Si). The last layer was made of either EVA or PVB.

The glass layer being impervious, the infiltration of moisture is done in the long run through the module sides. The contact of moisture with  the two types of encapsulants (EVA, PVB) yield to different processes, as shown in the following subsections. 

\subsection{ Poly Vinyl Butyral (PVB) used as encapsulant}
Previous studies demonstrated that, water slowly diffuse through the PVB as temperature \citep{baya28,baya29} increases, however the PVB polymer does not react with water molecules. Once the droplets of water reach the semiconductor in a form of humidity, a number of chemical reactions will occur due to the interaction between water and the semiconducting cells.  The water molecules initially dissociate according to the reaction \ref{re1}

\begin{equation}\label{re1}
H_{2}O \rightleftarrows H^{+}+OH^{-}
\end{equation}

The hydroxide ions OH$^{-}$ attach their hanging bonds at the TCO/Si interface and at the vacancies levels in the bulk to form silanols \cite{baya17,baya18}:

\begin{equation}\label{re2}
Si- + -OH \rightleftarrows Si-OH
\end{equation}

Other silanols are also formed at the TCO/Si interface by hydration of the $Si-H$ bonds resulting from the hydrogen passivation of the hanging bondings. This produces more H$^{+}$ ions as shown in equation \ref{re3}

\begin{equation} \label{re3}
Si-H+H_{2}O \rightleftarrows Si-OH+2H^{+}
\end{equation}

The hydrogen ions (H$^{+}$) resulting from the self-ionization of water molecules in equation \ref{re1} and those resulting from the hydration of the Si$-$H bonds as shown in reaction \ref{re3} mostly enters in the silicon interstices. However, some of those ions hydrogen react with silanols yielding to an amphoteric character as shown in reactions \ref{re4} and \ref{re5}

\begin{equation} \label{re4}
Si-OH+H^{+} \rightleftarrows Si-OHH^{+}
\end{equation}
\begin{equation}\label{re5}
Si-OH \rightleftarrows SiO^{-}+H^{+}
\end{equation}

\subsection{ Ethylene Vinyl Acetate (EVA) used as encapsulant}

Similarly to PVB, EVA is known to permit more water diffusion with a rate between 0.05 to 0.13 $\%$ \citep{baya28,baya29}. Additionaly, the EVA react with water molecules, so  the molecules of water  will react with the polymer to form acetic acid according to the reaction shown in reaction \ref{re6}


\begin{multline}\label{re6}
\{[-CH_{2}-CH(CH_{3}COO)-]+H_{2}O 
\rightleftarrows \\ [-CH_{2}-CH_{2}-]+CH_{3}COOH+ \frac{1}{2}O_{2} 
)
  \}
\end{multline}

The acid subsequently attacks metallic contacts and  to form metal acetates as described in reaction \ref{re7}

\begin{equation} \label{re7}
2nCH_{3}COOH+M_{m}O_{n} \longrightarrow M_{m}(CH_{3}COO)_{2n}+nH_{2}O
\end{equation}

Thus, at the level of the semiconducting material of PV cells, the molecules introduce metallic impurities in the bulk, in addition to the defects which occur in the case of PVB.

\begin{figure}[!htbp]
        \begin{center}
            \includegraphics[scale=0.25]{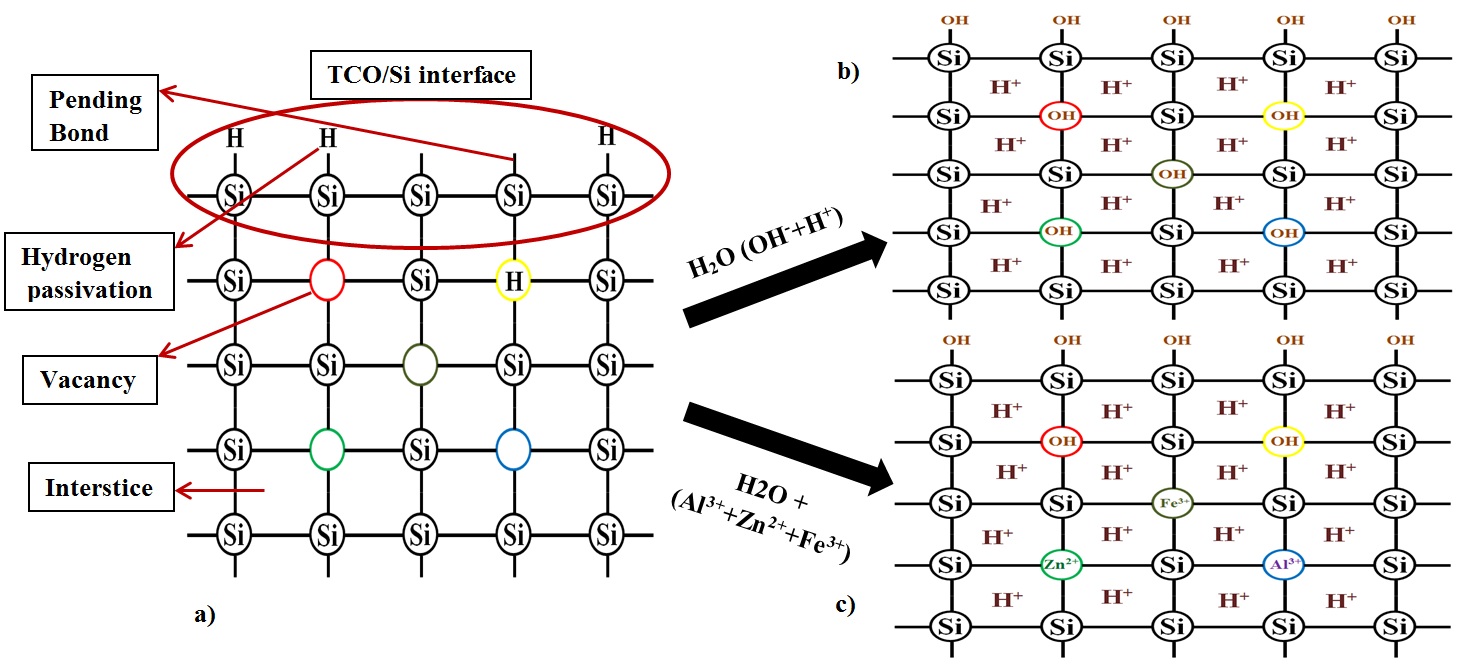}
        \end{center}
        \caption{Illustration of changing in the lattice structure. Initially, there are hydrogen and dangling bonds (a). Due to moisture, they become silanols and there are hydrogen ions which occur. These ions go to interstices (b). Some metallic ions can occur if  EVA  is as encapsulant and which stay in vacancies (c).}
        \label{figure2}
    \end{figure}

\begin{figure}[!htbp]
    \begin{center}
        \includegraphics[scale=0.25]{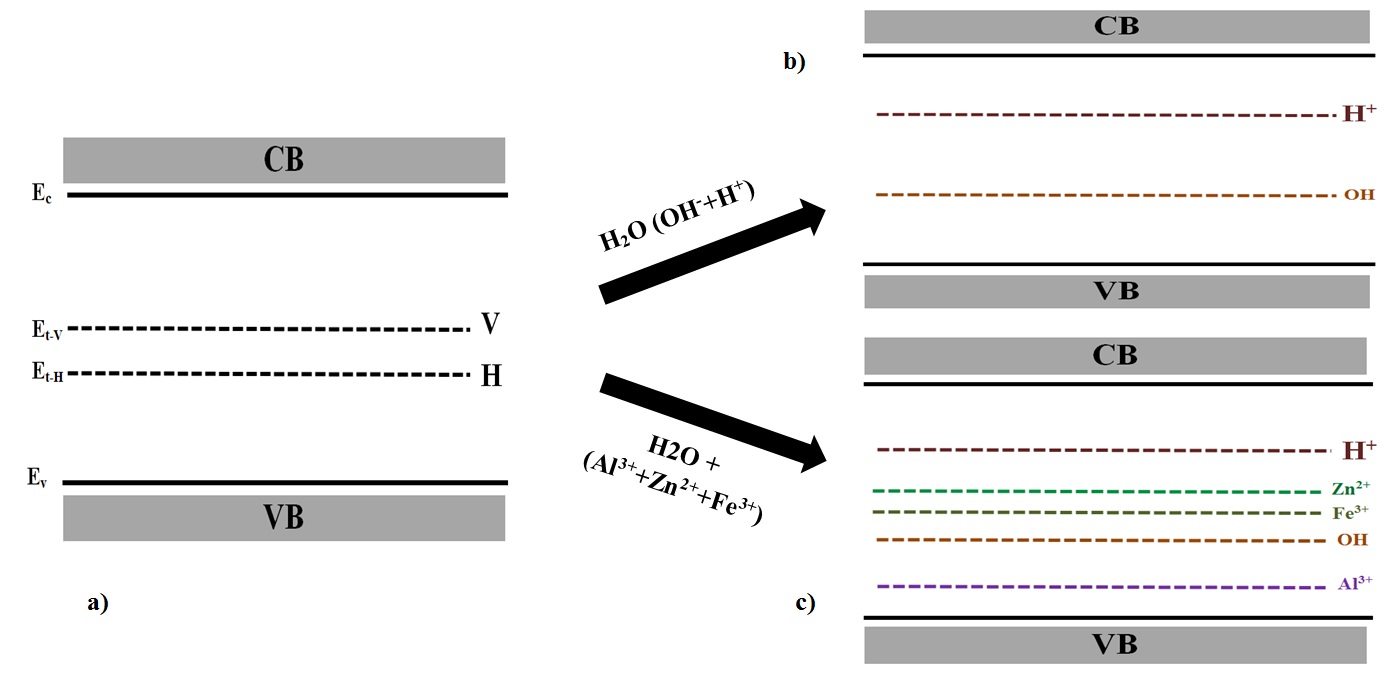}
    \end{center}
    \caption{Illustration of changing in the band diagram. Initially, there are some levels in the band gap due to hydrogen and hanging bonding (a). Due to moisture, those levels are replaced by a level corresponding to silanols and hydrogen ions in interstice (b). If EVA is used as an encapsulant, others levels are added in the band gap due to metallic ions in vacancies. (c)}
    \label{figure3}
\end{figure}

Subsequently, to the equation describes above, the model of figures \ref{figure2} and \ref{figure3} were used to further explain the chemical mechanism involved. In figure \ref{figure2} we modelled the different changes arising in the lattice structure of the material. In figure \ref{figure2}a we assume that  the surface there are hydrogen (Si$-$H) due to passivation, and some hanging bonds not passivated. Additionally, there could also be a presence of hydrogen, vacancies, and other minor defects from the bulk material. The reactions of equation \ref{re2} and \ref{re3} lead to a production of silanols as shown in figure \ref{re2}b, the Hydrogen ions from  reactions \ref{re2} and \ref{re3} will easily go to the interstice, since they have a smaller radius. 

When EVA is used as an encapsulant, metallic ions may be generated according to equation \ref{re7} as shown if figure \ref{figure2}c. The metallic ions, for instance, Al$^{3+}$ and Zn$^{2+}$, when aluminium is used as contact and ZnO as a donor layer respectively, will fill the vacancies according to their size. In a tropical area with a presence of dust and high occurrence of rainfall, the presence of  Fe$^{3+}$ should also be considered, given the high presence of these ions in the dust in those regions.

The modification of the lattice structure also leads to the modification of the band diagram as shown in figure \ref{figure3}.  Defects and impurities from previous reactions, introduce energies levels in the band gap. Namely, the energies levels due to hydrogen (Si$-$H) and dangling bonds (Si$-$) in figure \ref{figure3}a disappeared, and there are new energies levels as shown in figure \ref{figure3}b and \ref{figure3}c. In fact, in figure \ref{figure3}b, the levels due to Si$-$OH and H$^{+}$, replace the previous level due to  Si$-$H (hydrogen) and $Si-$ (dangling bonds). In figure \ref{figure3}c, the metallic ions which occur when using EVA (Al$^{3+}$ and Zn$^{2+}$), or from dust (Fe$^{3+}$), add more levels (corresponding to each ion) in the band gap.

 \section{Simulation details}

\begin{figure}[!htbp]
    \begin{center}
        \includegraphics[scale=0.3]{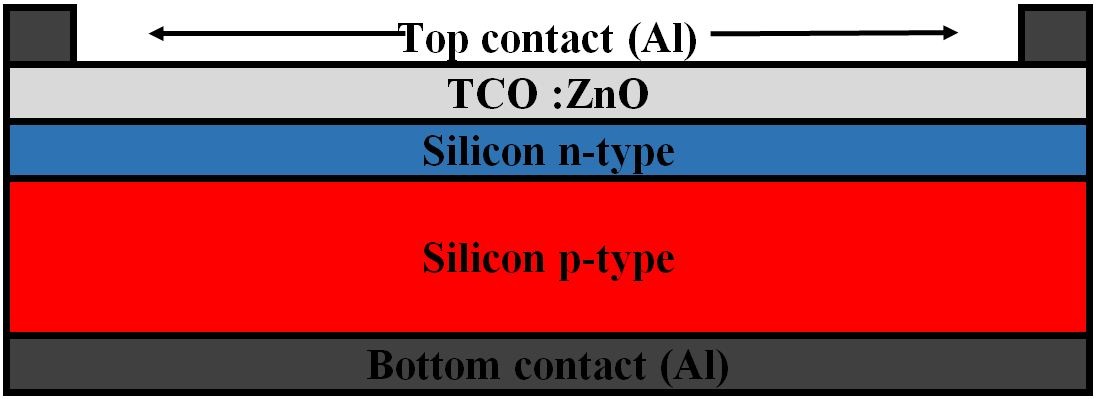}
    \end{center}
    \caption{Illustration of the model used for the PV simulation on SCAPS, a three layer structure was chosen with front and back contact in Aluminum.}
    \label{figure2a}
\end{figure}

The fast development of PV technology has also led to an increase in a number of simulation models and tools used to predict and approximate the operation of monocrystalline, polycrystalline and amorphous materials based solar cells. Amongst these models, the SCAPS simulator stands amongst those with  advanced algorithm taking into account the higher number of parameters \cite{baya19,baya20}.  We used as absorber a mono-crystalline $p$-doped silicon, whereas the buffer was taken to be a monocrystalline $n$-doped silicon and the window layer was made of ZnO, the contacts were made of aluminium which made a  non-rectifying junction with the semiconducting layers. The schematic of the structure used for this simulations is presented in figure \ref{figure2a}. This structure was input in the 3.3.00 version of the SCAPS software using the well-known convention, the light illumination was taken from the front contact, at a light power of one sun (1000W/m$^2$), with a transmission of $100$ \%.

Various parameters were used during this simulation, most parameters were obtained from the literature\cite{baya21,baya22}, with few parameters, reasonably estimated. the data of table \ref{table1} shows the parameters used for the doped silicon and the ZnO used throughout this simulation.

\begin{table*}[!htbp]
    \begin{center}
        \begin{tabular}{|l|c|c|c|}
            \hline
            & Si $p$-doped      & Si $n$-doped       &       i: ZnO \\
            \hline
            Thickness : $W$ $(\mu m)$                                 & $98 $               &  $2    $                 &    $   0.05$ \\
            \hline
            Profil                                              & Uniform            &        Uniform         &       Uniform \\
            \hline
            Properties                                           & Pure A (y=0)        &        Pure A (y=0)     &    Pure A (y=0)  \\
            \hline
            Gap energy : $E_{g}$ $(eV)$                           & $1.12  $             &  $1.12 $                  &     $  3.3$ \\
            \hline
            Electron afinity : $\chi$ $(eV)$                     & $4.05$               & $ 4.05$                   &     $  4.45$ \\
            \hline
            Relative permitivity  : $\epsilon_{r}$                    & $11.9  $             & $ 11.9 $                  &      $ 9$ \\
            \hline
            Density of state in $CB$ : $N_{C}$ $(cm^{-3})$   & $2.8\times 10^{19}$ & $2.8\times 10^{19} $  &   $  2.2\times 10^{18}$ \\
            \hline
            Density of state in $VB$ : $N_{V}$ $(cm^{-3})$   & $1.04\times 10^{17}$ & $1.04\times 10^{17} $ & $  1.8\times 10^{19}$ \\
            \hline
            Thermale velocity of electrons :$V_{thn}$ $(cm/s)$       &$ 10^{7} $        &  $ 10^{7}  $              &   $   10^{7} $\\
            \hline
            Thermale velocity of holes :$V_{thp}$ $(cm/s)$           &$ 10^{7}  $           &  $ 10^{7}  $          &   $   10^{7}$ \\
            \hline
            Electrons mobility : $\mu_{n}$ $(cm^{2}/V_{s})$       & $1500  $             &  $  1500$                 &    $  10^{2} $\\
            \hline
            Holes mobility : $\mu_{p}$ $(cm^{2}/V_{s})$           &$ 450     $           &   $ 450 $                 &    $   25$ \\
            \hline
            Density of donnors : $N_{D}$ $(cm^{-3})$           &$   10^{1}   $          &   $ 10^{16} $               &    $ 10^{18}$ \\
            \hline
            Density of acceptors : $N_{A}$ $(cm^{-3})$         &$ 10^{15}     $           &   $ 10^{1} $             &    $ 10^{1}$ \\
            \hline
            Coefficient of absorption                      &  SCAPS (Si.abs)      &   SCAPS (Si.abs)              &   SCAPS (ZnO.abs) \\
            \hline

        \end{tabular}
        \caption{Recapitulation table of parameters used for the simulation for each layers. From left to right, denomination of the parameters, $p$-doped silicon layer, $n$-doped silicon layer, zinc oxide layer. Units are given in the first column}
        \label{table1}
    \end{center}
\end{table*}

According to our moisture modelling explained in figures \ref{figure2} and \ref{figure3}, three situations were simulated: Without moisture, with moisture and with moisture containing metallic ions.

First of all, we simulate the situation without moisture. In this case, as explained above, we assumed that we have two defects: Si$-$H (which was hydrogen bound) and Si$-$ (which was vacancy). These defects were registered as neutral with a uniform density. The capture cross-section of electrons and holes were respectively 1.0 $\times$10$^{-17}$ cm$^2$ and 1.0 $\times$10$^{-18}$ cm$^2$ for Si$-$H; and 1.0 $\times$10$^{-16}$ cm$^2$ and 1.0 $\times$10$^{-14}$ cm$^2$ for Si$-$ \cite{baya21,baya22}. The activation energy for Si$-$H was 0.39 eV \cite{baya21,baya22}. For  Si$-$, the activation energy was 0.47 eV \cite{baya21,baya22}. The values of the densities used are recorded in table \ref{table2}.

\begin{table}[!htbp]
    \begin{center}
        \begin{tabular}{|c|c|c|c|}
            \hline
            & ZnO/Si  & Si $n$-doped & Si $p$-doped \\
              & $(cm^{-2})$     & $(cm^{-3})$ &   $(cm^{-3})$  \\
            \hline
            Si$-$H &  1.0 $\times$10$^{11}$     & 1.0 $\times$10$^{12}$  &   4.9 $\times$10$^{13}$ \\
            \hline
            Si$-$ &  1.0 $\times$10$^{4}$     & 1.0 $\times$10$^{4}$  &   4.9 $\times$10$^{5}$ \\
            \hline
        \end{tabular}
        \caption{Densities of the defects (Si$-$H and Si$-$) in the cell without moisture. From left to the right are the density at the ZnO/Si interface, in the n-doped region and the p-doped region.}
        \label{table2}
    \end{center}
\end{table}

Secondly, we simulated the situation related to moisture ingress while using an encapsulant like PVB, which does not react with water. According to  reactions \ref{re2} and \ref{re3}, the previous defects (Si$-$H and Si$-$) were replaced by Silanol (Si$-$OH) and (H$^{+}$) as shown in figure \ref{figure2}. during the simulation, the (H$^{+}$) was introduced into the bulks regions as a single donor. The capture cross section of electrons and holes were 1.0 $\times$10$^{-14}$ cm$^2$ and 1.0 $\times$10$^{-16}$ cm$^2$ respectively \cite{baya21,baya22}. The activation energy was 0.68 eV with respect to the highest level of the Valence Band \cite{baya21,baya22}. Silanols (Si$-$OH)  introduced at the interface  of ZnO/Si as a neutral defect, and as an amphoteric defect in the bulk material according to the reactions \ref{re4} and \ref{re5}. The activation energy was 0.25 eV with respect to the highest level of the Valence Band \cite{baya21,baya22} at the interface ZnO/Si. The captures cross section of electrons and holes were  3.0 $\times$10$^{-17}$ cm$^2$ and 3.0 $\times$10$^{-15}$ cm$^2$ respectively \cite{baya21,baya22} at the interface ZnO/Si, similar parameters were also used in the bulk material.
Throughout the simulation a linear gradient was used to characterized the densities of defect as shown in table  \ref{table3}

\begin{table*}[!htbp]
    \begin{center}
        \begin{tabular}{|c|c|c|c|c|c|}
            \hline
            & ZnO/Si  & \multicolumn{2}{c|}{Si $n$-doped} & \multicolumn{2}{c|}{Si $p$-doped} \\
             &  $(cm^{-2})$ & \multicolumn{2}{c|}{$(cm^{-3})$}  & \multicolumn{2}{c|}{$(cm^{-3})$}   \\
            \cline{3-6}
                  &             & $x=0$ $\mu m$      &      $x=2$ $\mu m$   &  $x=0$ $\mu m$  & $x=98$ $\mu m$ \\
            \hline
            Si$-$OH       & 1.0 $\times$10$^{15}$  & 1.0 $\times$10$^{11}$  &    1.0 $\times$10$^{10}$   & 1.0 $\times$10$^{10}$ & 1.0 $\times$10$^{2}$  \\
            \hline
            H$^{+}$      & 0  & 2.0 $\times$10$^{14}$   &  2.0 $\times$10$^{12}$  & 2.0 $\times$10$^{12}$ & 1.0 $\times$10$^{2}$ \\
            \hline
        \end{tabular}
        \caption{ Densities of silanols (Si$-$OH) and hydrogen ions (H$^{+}$) in the cell with moisture while using PVB as encapsulant. The density of silanol and hydrogen ion are varied in the different layers.}
        \label{table3}
    \end{center}
\end{table*}


The case of moisture ingress in PV encapsulated with EVA was simulated. In this simulation case, besides the (Si$-$OH and H$^{+}$), metallic ions such as Al$^{3+}$; Zn$^{2+}$ and Fe$^{3+}$ were added to the simulation profile as described in chemical consideration section.  From the simulation of cells where the PVB encapsulant was used, were kept at their positions at different interfaces using the densities previously used.

 The Zn$^{2+}$ was introduced as a double donor withanactivation energy of 0.55 eV . The captures cross section of electrons and holes were  1.0 $\times$10$^{-13}$ cm$^2$ and 1.0 $\times$10$^{-15}$ cm$^2$ respectively for  the level $+/0$, and 1.0 $\times$10$^{-12}$ cm$^2$ and 1.0 $\times$10$^{-16}$ cm$^2$ for the level $2+/+$ .  Al$^{3+}$ was introduced as a multilevel defect with an activation energy of 0.063 eV. A constant capture cross section was also used for different levels of energy varying from  1.0 $\times$10$^{-19}$ cm$^2$ to   1.0 $\times$10$^{-13}$ cm$^2$.   The Fe$^{3+}$ was introduced as a multilevel defect with an activation energy of 0.10 eV. For this element the capture cross section were also varied from 1.0 $\times$10$^{-18}$ cm$^2$ to 1.0 $\times$10$^{-10}$ cm$^2$. During this simulations, a linear gradient of densities as describes in table \ref{table4} were used. 

\begin{table*}[!htbp]
    \begin{center}
        \begin{tabular}{|c|c|c|c|c|}
            \hline
            & \multicolumn{2}{c|}{Si $n$-doped} & \multicolumn{2}{c|}{Si $p$-doped} \\
            & \multicolumn{2}{c|}{$(cm^{-3})$}  & \multicolumn{2}{c|}{$(cm^{-3})$}   \\
            \cline{2-5}
            & $x=0$ $\mu m$      &      $x=2$ $\mu m$   &  $x=0$ $\mu m$  & $x=98$ $\mu m$ \\
            \hline
            Zn$^{2+}$         & 1.0 $\times$10$^{16}$  &    1.0 $\times$10$^{14}$   & 1.0 $\times$10$^{14}$ & 1.0 $\times$10$^{2}$  \\
            \hline
            Al$^{3+}$      & 1.0 $\times$10$^{16}$  &    1.0 $\times$10$^{14}$   & 1.0 $\times$10$^{14}$ & 1.0 $\times$10$^{2}$  \\
            \hline
            Fe$^{3+}$       & 1.0 $\times$10$^{16}$  &    1.0 $\times$10$^{14}$   & 1.0 $\times$10$^{14}$ & 1.0 $\times$10$^{2}$  \\
            \hline
        \end{tabular}
        \caption{ Densities of metallic ions (Al$^{3+}$, Zn$^{2+}$ and Fe$^{3+}$) in the bulk when the cell has EVA as encapsulant. The density of each ion decreases from 1.0 $\times$10$^{16}$ cm$^{-3}$ to 1.0 $\times$10$^{14}$ cm$^{-3}$ in the layer Si $n$-doped, and from 1.0 $\times$10$^{14}$ cm$^{-3}$ to 1.0 $\times$10$^{2}$ cm$^{-3}$ in the layer Si $p$-doped.}
        \label{table4}
    \end{center}
\end{table*}

  We simulate at 300 K, with a constant series resistances of 0.1 $\Omega$.cm$^{2}$ and shunt resistance of 1.0 $\times$10$^{30}$ $\Omega$.cm$^{2}$. Throughout this simulation, the radiative and Auger recombination was not taken into account. But the Shockley-Read-Hall (SRH) recombinations were accounted for, using the model of PV impurity in  \cite{baya23}.

\section{Results and Discussion}

   \subsection{Effects of moisture on Spectral Response}

\begin{figure}[!htbp]
    \begin{center}
        \includegraphics[scale=0.6]{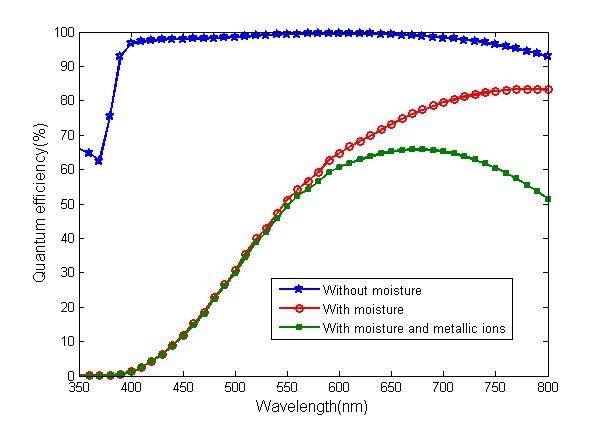}
    \end{center}
    \caption{Variation of quantum efficiency with the light wavelength.  The graph for the PV operating without moisture is plotted in blue (top plot with starred dot). The graph for the PV with moisture is plotted in shades of red (circular dot). The case for PV with moisture and metallic ions in shade of green (squared dot).}
    \label{figure5}
\end{figure}

In figure \ref{figure5} we showed the spectral responses, describing the evolution of quantum efficiency with the varying wavelength for solar cells without moisture, with moisture, and with moisture containing metallic ions. 
 
\begin{table*}[!htbp]
    \begin{center}
        \begin{tabular}{|c|c|c|c|c|c|c|c|}
            \hline
            & V$_{OC}$   & J$_{SC}$& FF & PCE & J$_{MP}$ & V$_{MP}$ & P$_{max}$ \\
            & (V)    & (mA.cm$^{-2}$) &   $(\%)$ & $(\%)$ & (mA.cm$^{-2}$) & $(V)$ & (mW.cm$^{-2}$) \\
            \hline
            without moisture &  0.67     & 34.17 &   82.80  &18.57  & 32.59&0.57 & 18.56\\
            \hline
            with moisture &  0.62     & 22.56  &   81.78 & 11.46 & 21.34 & 0.54 & 11.45\\
            \hline
            with moisture and metallic ions &  0.59     & 15.17  &   79.20 & 7.04 &14.09  & 0.50 &7.04 \\
            \hline
        \end{tabular}
        \caption{Photovoltaic parameters of cells when used without moisture, with moisture, and with moisture containing metallic ions.  V$_{OC}$ is the Open Circuit Voltage, J$_{SC}$ is the Density of current in  Circuit, FF is the Fill Factor, PCE is the Power Conversion Efficiency. J$_{MP}$, V$_{MP}$ and P$_{max}$ are the density of current, the voltage and the maximalpower at the functioning  point respectively.}
        \label{table5}
    \end{center}
\end{table*}
 
 Figure \ref{figure5} shows a net loss of quantum efficiency (QE) when the moisture ingress in the semiconducting material. The losses are increase when there is a presence of metallic ions in the moisture. This leads to a significant loss of electrical performance and power output during the operation. This could be justify by the fact that, by modeling the moisture as defects and impurities in the semiconducting material of (figure \ref{figure2}), there is an introduction of energy levels in the band gap as described in (figure \ref{figure3}), the moisture  henceforth creates several traps and recombination centers which lead to the modification of the properties of the cells. The traps and recombination centers reduce the diffusion length and the carriers lifetime, hence the losses in quantum efficiency. Those results were corroborated by the study done by Babaee et Al \cite{baya24} on the popular simulation software SILVACO with monocrystalline silicon.

 The top curve of figure  \ref{figure5} shows the spectral response of a monocrystalline silicon PV cell, with a noticeable decrease from around $\lambda=$650 nm due to higher reflectivity and an increase in surface recombination.  A sharp decrease is also noticed around the radiation with  $\lambda=$400 nm. This observation is due to the fact that, silicon is a semiconductor with an indirect band-gap, therefore short wavelengths are absorbed mostly in shallow depths and long wavelengths mostly enter more in-depth before their absorption \cite{baya26}.  We noticed a significant change in the spectral profile, as the moisture goes into the cells. Below $\lambda=$400 nm, there is no detectable radiation, a stiff increase is observed from  400 nm to 600 nm, while the QE for the cells with moisture continue  to increase to reach their maximum around $\lambda=$800 nm in our study. For the case with metallic ions a maximum is reach around $\lambda=$650 nm, then the decrease of QE is further observed. Thus, there is a net lost on QE between the three states moisture free, with moisture, and moisture with metallic ions. This would translate into a significant loss of electrical efficiency or power when the PVs are operated with moisture in general and with a presence of metallic ions in particular.

\subsection{Effects of moisture on IV}

\begin{figure}[!htbp]
    \begin{center}
        \includegraphics[scale=0.6]{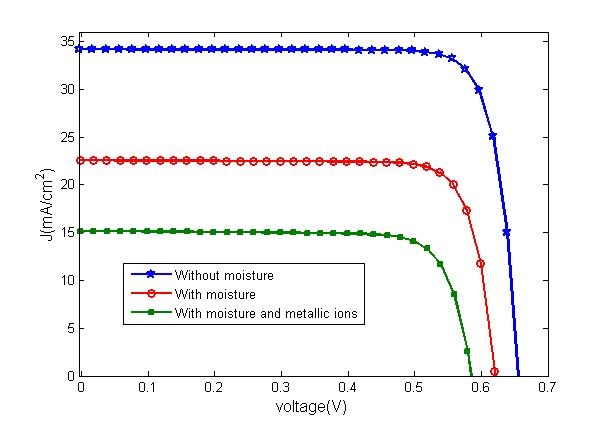}
    \end{center}
    \caption{Current-Voltage characteristics  in light conditions of  the cells for different situation: without moisture (top curve), with moisture (middle curve) and with moisture and metallic ions (lower curve).}
    \label{figure6}
\end{figure}

Figure \ref{figure6}  shows the variation of current density with  voltage under light condition.  The top plot correspond to the case without moisture, the middle plot being for the case with moisture and the lower plot for  cells with moisture and the presence of metallic ions. Key parameters found during this study are presented in table \ref{table5}.

The curves on figure \ref{figure6} shows that, from the curve of a moisture free cells, there is a drop of almost 30 $\%$ in the current density when there is moisture in the cells, and that decrease is more than  50 $\%$ when we have a presence of metallic ions in the moisture.

\begin{figure*}[!htbp]
\centering
   \begin{subfigure}{.5\textwidth}
  \centering
  \includegraphics[width=1\linewidth]{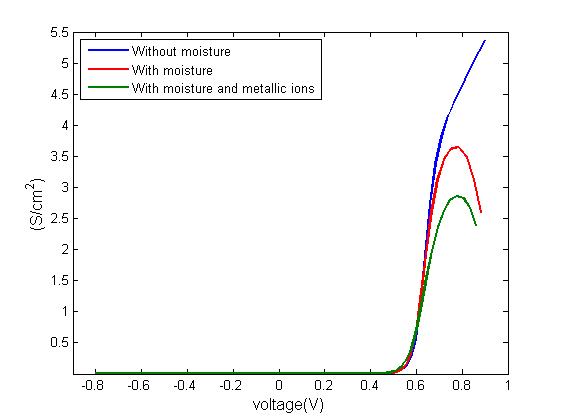}
  \caption{}
  \label{figure7b}
\end{subfigure}%
  \begin{subfigure}{.5\textwidth}
  \centering
  \includegraphics[width=1\linewidth]{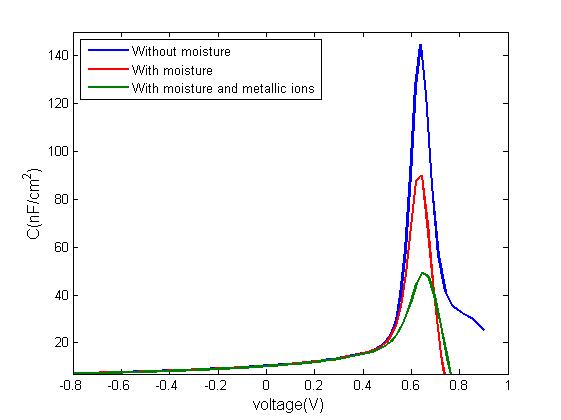}
  \caption{}
  \label{figure7c}
\end{subfigure}%

\caption{ Dependence of conductivity and capacitance of the solar cells with the increasing voltages. The top plot describe the cells without moisture (blue); the middle plot the one with moisture (red); and the lower plot the cells with moisture and metallic ions (green)}
\label{figure7} 
\end{figure*}

The change of capacitance and conductivity with the voltage in figures \ref{figure7c} and \ref{figure7b} respectively emphasized the changes that could occurred in the material while being operated in a humid environment with the presence of metallic ions. The curves also imply an increase of the series resistances which are due to the modification of contact at the interface between the materials and the metallic contacts.  This could be justified by using the sphere model, which stipulate that both the carriers and the moisture could be considered as spheres \cite{baya30}. Therefore, it will be harder for the carriers to move inside the network due to the presence of moisture presented here as other spheres. The "slow-down" of carriers lead to the increase of series resistance and the ideality factors. That yield to the decrease in fill factor as shown in table \ref{table5}, so it is reasonable to say that, the presence of moisture and metallic ions reduce the FF, hence the performance of the PV module. These results are corroborated by the experimental work done by Weerasinghe et Al \cite{baya27}.

We further demonstrated with the results in table \ref{table5} that there is a drop on short circuit current J$_{SC}$ and power conversion efficiency PCE, this is possibly due to the decrease in the generated and collected photo-current related to moisture ingress in the cell. That was due to the creations of traps and recombination centers as mentioned earlier, those phenomena usually reduce the carriers diffusion length, hence the drop in the value of these parameters. The drops in those parameters ultimately lead to the deterioration of the functioning point and the decrease of maximal power as shown in table \ref{table5}.

Table \ref{table5} also shows  the decrease on Open Circuit Voltage (V$_{oc}$). In fact, during the traps and recombination process there is a net reduction in the quantities of carriers in the Cells's network. Therefore, V$_{oc}$ decreases. However, previous study \cite{baya27} did not notice a significant change in open circuit voltage related to moisture ingress in the cells in the case of thin-film PV. 

In overall it could be said that metallic ions potentially add more traps and recombination in the semiconducting materials, therefore leading to more deterioration of the cells. So, it could be said that the decrease of all photovoltaic parameters are enhanced by the presence of metallic ions as shown throughout this work. 

\section{Conclusion}

In this paper, the influence of moisture on the operation of a mono-crystalline based silicon photovoltaic cells was studied using SCAPS software. The moisture was modelled by defects and impurities generated from the reaction between the water and silicon. So, silanol Si$-$OH  and hydrogen ions  H$^{+}$ were used as impurities. Further investigation was performed while metallic impurities such as (Zn$^{2+}$, Al$^{3+}$, Fe$^{3+}$) coming from dust and other elements such as encapsulant, TCO and metallic contacts. It was established from a chemical consideration that the degradation of the cells due to moisture also depend on the type of encapsulant as the use of EVA or PVB  gave different degradation processes.  

By analyses of the three cases where the PV module was operating moisture free; with moisture; and with moisture having metallic ions, it was established that there is a significant loss or reduction on the PV parameters with the apparition of moisture and metallic ions. For instance, a power loss of more than 50 \% was observed. This work has proven to be useful on predicting the degradation of a PV module based on the type of encapsulant used, therefore, these results could help on choosing the types of encapsulant more suitable for the tropical, humid and rainy environment. Namely, this work can also be useful for the production of efficient cells in wet regions, by helping  in the choice of the encapsulant that minimises the production of metallic ions. 

It is understood that further study in this direction using the most widely used type of Cells such as CIGS, CZTS, and organic cells will also help to better understand the operation and degradation mechanism of those type of Cells while use in the wet and humid environment.

\section*{Acknowledgements}

The Authors acknowleged Martial Watio KEYAMPI for it linguistic support during the initial phase of this manuscript. They also acknowledge Dr. Moise Dongho Nguimdo for the constructive discussion and assistance while writing this manuscript.

\section*{References}

\bibliography{Bayawa_et_al}

\begin{thebibliography}{10}
\expandafter\ifx\csname url\endcsname\relax
  \def\url#1{\texttt{#1}}\fi
\expandafter\ifx\csname urlprefix\endcsname\relax\def\urlprefix{URL }\fi
\expandafter\ifx\csname href\endcsname\relax
  \def\href#1#2{#2} \def\path#1{#1}\fi

\bibitem{baya1}
X.~Zhang, X.~Zhao, S.~Smith, J.~Xu, X.~Yu, Review of r\&d progress and
  practical application of the solar photovoltaic/thermal (pv/t) technologies,
  Renewable and Sustainable Energy Reviews 16~(1) (2012) 599 -- 617.
\newblock \href
  {http://dx.doi.org/http://dx.doi.org/10.1016/j.rser.2011.08.026}
  {\path{doi:http://dx.doi.org/10.1016/j.rser.2011.08.026}}.

\bibitem{baya2}
L.~E. Chaar, L.~lamont, N.~E. Zein, Review of photovoltaic technologies,
  Renewable and Sustainable Energy Reviews 15~(5) (2011) 2165 -- 2175.
\newblock \href
  {http://dx.doi.org/http://dx.doi.org/10.1016/j.rser.2011.01.004}
  {\path{doi:http://dx.doi.org/10.1016/j.rser.2011.01.004}}.

\bibitem{baya3}
O.~O. Ogbomo, E.~H. Amalu, N.~Ekere, P.~Olagbegi, A review of photovoltaic
  module technologies for increased performance in tropical climate, Renewable
  and Sustainable Energy Reviews 75 (2017) 1225 -- 1238.
\newblock \href
  {http://dx.doi.org/http://dx.doi.org/10.1016/j.rser.2016.11.109}
  {\path{doi:http://dx.doi.org/10.1016/j.rser.2016.11.109}}.

\bibitem{baya4}
K.~Branker, M.~Pathak, J.~Pearce, A review of solar photovoltaic levelized cost
  of electricity, Renewable and Sustainable Energy Reviews 15~(9) (2011) 4470
  -- 4482.
\newblock \href
  {http://dx.doi.org/http://dx.doi.org/10.1016/j.rser.2011.07.104}
  {\path{doi:http://dx.doi.org/10.1016/j.rser.2011.07.104}}.

\bibitem{baya5}
S.~Hegedus, A.~Luque, Achievements and Challenges of Solar Electricity from
  Photovoltaics, John Wiley \& Sons, Ltd, 2011, pp. 1--38.
\newblock \href {http://dx.doi.org/10.1002/9780470974704.ch1}
  {\path{doi:10.1002/9780470974704.ch1}}.

\bibitem{baya6}
N.~Park, C.~Han, D.~Kim, Effect of moisture condensation on long-term
  reliability of crystalline silicon photovoltaic modules, Microelectronics
  Reliability 53~(12) (2013) 1922 -- 1926.
\newblock \href
  {http://dx.doi.org/http://dx.doi.org/10.1016/j.microrel.2013.05.004}
  {\path{doi:http://dx.doi.org/10.1016/j.microrel.2013.05.004}}.

\bibitem{baya7}
M.~D. Kempe, A.~A. Dameron, M.~O. Reese, Evaluation of moisture ingress from
  the perimeter of photovoltaic modules, Progress in Photovoltaics: Research
  and Applications 22~(11) (2014) 1159--1171.
\newblock \href {http://dx.doi.org/10.1002/pip.2374}
  {\path{doi:10.1002/pip.2374}}.

\bibitem{baya13}
M.~Burgelman, P.~Nollet, S.~Degrave, Modelling polycrystalline semiconductor
  solar cells, Thin Solid Films 361 (2000) 527 -- 532.
\newblock \href
  {http://dx.doi.org/http://dx.doi.org/10.1016/S0040-6090(99)00825-1}
  {\path{doi:http://dx.doi.org/10.1016/S0040-6090(99)00825-1}}.

\bibitem{baya8}
S.~Mekhilef, R.~Saidur, M.~Kamalisarvestani, Effect of dust, humidity and air
  velocity on efficiency of photovoltaic cells, Renewable and Sustainable
  Energy Reviews 16~(5) (2012) 2920 -- 2925.
\newblock \href
  {http://dx.doi.org/http://dx.doi.org/10.1016/j.rser.2012.02.012}
  {\path{doi:http://dx.doi.org/10.1016/j.rser.2012.02.012}}.

\bibitem{baya9}
H.~C. Weerasinghe, S.~E. Watkins, N.~Duffy, D.~J. Jones, A.~D. Scully,
  Influence of moisture out-gassing from encapsulant materials on the lifetime
  of organic solar cells, Solar Energy Materials and Solar Cells 132 (2015) 485
  -- 491.
\newblock \href
  {http://dx.doi.org/http://dx.doi.org/10.1016/j.solmat.2014.09.030}
  {\path{doi:http://dx.doi.org/10.1016/j.solmat.2014.09.030}}.

\bibitem{baya10}
T.~Kim, N.~Park, D.~Kim, The effect of moisture on the degradation mechanism of
  multi-crystalline silicon photovoltaic module, Microelectronics Reliability
  53~(9) (2013) 1823 -- 1827, european Symposium on Reliability of Electron
  Devices, Failure Physics and Analysis.
\newblock \href
  {http://dx.doi.org/http://dx.doi.org/10.1016/j.microrel.2013.07.047}
  {\path{doi:http://dx.doi.org/10.1016/j.microrel.2013.07.047}}.

\bibitem{baya11}
K.~Morita, T.~Inoue, H.~Kato, I.~Tsuda, Y.~Hishikawa, Degradation factor
  analysis of crystalline-si pv modules through long-term field exposure test,
  in: 3rd World Conference onPhotovoltaic Energy Conversion, 2003. Proceedings
  of, Vol.~2, 2003, pp. 1948--1951 Vol.2.

\bibitem{baya14}
O.~Simya, A.~Mahaboobbatcha, K.~Balachander, A comparative study on the
  performance of kesterite based thin film solar cells using scaps simulation
  program, Superlattices and Microstructures 82 (2015) 248 -- 261.
\newblock \href
  {http://dx.doi.org/http://dx.doi.org/10.1016/j.spmi.2015.02.020}
  {\path{doi:http://dx.doi.org/10.1016/j.spmi.2015.02.020}}.

\bibitem{baya15}
N.~Khoshsirat, N.~A.~M. Yunus, Numerical simulation of cigs thin film solar
  cells using scaps-1d, in: 2013 IEEE Conference on Sustainable Utilization and
  Development in Engineering and Technology (CSUDET), 2013, pp. 63--67.
\newblock \href {http://dx.doi.org/10.1109/CSUDET.2013.6670987}
  {\path{doi:10.1109/CSUDET.2013.6670987}}.

\bibitem{baya28}
R.~A. Thomas~Swonke, Impact of moisture on pv module encapsulants (2009).
\newblock \href {http://dx.doi.org/10.1117/12.825943}
  {\path{doi:10.1117/12.825943}}.

\bibitem{baya29}
N.~Kim, C.~Han, Experimental characterization and simulation of water vapor
  diffusion through various encapsulants used in pv modules, Solar Energy
  Materials and Solar Cells 116~(Supplement C) (2013) 68 -- 75.
\newblock \href
  {http://dx.doi.org/https://doi.org/10.1016/j.solmat.2013.04.007}
  {\path{doi:https://doi.org/10.1016/j.solmat.2013.04.007}}.

\bibitem{baya17}
P.~Allongue, C.~H. de~Villeneuve, S.~Morin, R.~Boukherroub, D.~D. Wayner, The
  preparation of flat h–si(111) surfaces in 40
  Electrochimica Acta 45~(28) (2000) 4591 -- 4598.
\newblock \href
  {http://dx.doi.org/https://doi.org/10.1016/S0013-4686(00)00610-1}
  {\path{doi:https://doi.org/10.1016/S0013-4686(00)00610-1}}.

\bibitem{baya18}
N.~B. Ahmed, O.~Ronsin, L.~Mouton, C.~Sicard, C.~Yepremian, T.~Baumberger,
  R.~Brayner, T.~Coradin, The physics and chemistry of silica-in-silicates
  nanocomposite hydrogels and their phycocompatibility, J. Mater. Chem. B 5
  (2017) 2931--2940.
\newblock \href {http://dx.doi.org/10.1039/C7TB00341B}
  {\path{doi:10.1039/C7TB00341B}}.

\bibitem{baya19}
J.~Verschraegen, M.~Burgelman, Numerical modeling of intra-band tunneling for
  heterojunction solar cells in scaps, Thin Solid Films 515~(15) (2007) 6276 --
  6279, proceedings of Sympodium O on Thin Film Chalcogenide Photovoltaic
  Materials, EMRS 2006 Conference.
\newblock \href {http://dx.doi.org/https://doi.org/10.1016/j.tsf.2006.12.049}
  {\path{doi:https://doi.org/10.1016/j.tsf.2006.12.049}}.

\bibitem{baya20}
K.~Decock, S.~Khelifi, M.~Burgelman, Modelling multivalent defects in thin film
  solar cells, Thin Solid Films 519~(21) (2011) 7481 -- 7484, proceedings of
  the EMRS 2010 Spring Meeting Symposium M: Thin Film Chalcogenide Photovoltaic
  Materials.
\newblock \href {http://dx.doi.org/https://doi.org/10.1016/j.tsf.2010.12.039}
  {\path{doi:https://doi.org/10.1016/j.tsf.2010.12.039}}.

\bibitem{baya21}
M.~Otfried, Semiconductors, Springer, 2012.

\bibitem{baya22}
M.~Levinshtein, S.~Rumyantsev, M.~Shur, Handbook Serie on : semiconductor
  parameter, World Scientific, 1996.

\bibitem{baya23}
M.~J. Keevers, M.~A. Green, Efficiency improvements of silicon solar cells by
  the impurity photovoltaic effect., J. Applied Physics 75 (1994) 4022 -- 4031.
\newblock \href {http://dx.doi.org/http://dx.doi.org/10.1063/1.356025}
  {\path{doi:http://dx.doi.org/10.1063/1.356025}}.

\bibitem{baya24}
S.~Babaee, S.~Ghozati, The study of 1 mev electron irradiation induced defects
  in n-type and p-type monocrystalline silicon, Radiation Physics and Chemistry
  141 (2017) 98 -- 102.
\newblock \href
  {http://dx.doi.org/http://dx.doi.org/10.1016/j.radphyschem.2017.06.012}
  {\path{doi:http://dx.doi.org/10.1016/j.radphyschem.2017.06.012}}.

\bibitem{baya26}
S.~M. Sze, Semiconductor devices, New york: John Wiley, 1981.

\bibitem{baya30}
Theory of simple liquids, in: J.-P. Hansen, , I.~R. McDonald (Eds.), Theory of
  Simple Liquids (Fourth Edition), fourth edition Edition, Academic Press,
  Oxford, 2013, pp. i --.
\newblock \href
  {http://dx.doi.org/https://doi.org/10.1016/B978-0-12-387032-2.00013-1}
  {\path{doi:https://doi.org/10.1016/B978-0-12-387032-2.00013-1}}.

\bibitem{baya27}
H.~C. Weerasinghe, S.~E. Watkins, N.~Duffy, D.~J. Jones, A.~D. Scully,
  Influence of moisture out-gassing from encapsulant materials on the lifetime
  of organic solar cells, Solar Energy Materials and Solar Cells 132 (2015) 485
  -- 491.
\newblock \href
  {http://dx.doi.org/http://dx.doi.org/10.1016/j.solmat.2014.09.030}
  {\path{doi:http://dx.doi.org/10.1016/j.solmat.2014.09.030}}.

\end{thebibliography}

\end{document}